\documentstyle[multicol,aps,epsf]{revtex}
\begin{document}
\title{Subsonic ion-acoustic solitons}
\author{H. Abbasi$^{a,b}$\footnote{abbasi@ipm.ir}, H. Hakimi Pajouh$^{b}$, and H. A. Shah$^{c}$}
\address{$^a$Faculty of Physics, Amir Kabir University of Technology,
P. O. Box 15875-4413, Tehran, Iran \\
$^b$Institute for Studies in Theoretical Physics and Mathematics,
P. O. Box 19395-5531, Tehran, Iran\\
$^c$ Department of Physics, Government College University, Lahore 54000, Pakistan 
}
\maketitle

\begin{abstract}
In this paper, the nonlinear theory of plasma waves is extended to the plasmas that their equilibrium state are specified by the non-Maxwellian (here kappa) distribution. We believe that the extension is very important since most of the space and some of laboratory plasmas are not in the Maxwellian equilibrium. Although the linear theory of this issue has been known for the decades but, to our knowledge, this is the first attempt in opening the gate to the nonlinear world of plasma waves with the non-Maxwellian equilibrium. As an example the ion-acoustic solitons are studied in this framework taking into account the electron trapping in the trough of longitudinal field. It is shown, as the most important result regarding to the non-Maxwillian equilibrium, that there is the possibility for the ion-acoustic solitons to move subsonically. The solitons velocity and their width are monotonically increasing functions of spectral index ($\kappa$) and approach to the their Maxwellian values as $\kappa \rightarrow \infty$.
\end{abstract}

\section{Introduction}

In the natural space environment, e.g., planetary magnetospheres, astrophysical plasmas, and the solar wind, plasmas are generally observed to possess a non-Maxwellian high energy tail \cite{1,2,3,4,5}. A useful distribution function to model such plasmas is the generalized Lorentzian (kappa) distribution. Important features of $\kappa$ distribution are that, first, at high velocities the distribution obeys an inverse power law, and that, second, for all velocities, in the limit as the spectral index approaches to large values the distribution tends to the Maxwellian distribution. In this sense, the kappa distribution is a generalization of the Maxwellian distribution. Vasyliunas appears to have been the first to employ the general form of the kappa distribution and to note its relation to the Maxwellian \cite{1}. Kappa distribution have been used to analyze and interpret spacecraft data on the earth's magnetospheric plasma sheet \cite{2}, the solar wind \cite{3}, Jupiter \cite{4}, and Saturn \cite{5}. In practice it is found that many space plasmas can be modeled more effectively by a superposition of kappa distribution than by Maxwellians. In the context of both space and laboratory plasmas Hasegawa {\it et.al.} \cite{6} showed that the equilibrium state of the distribution function for a plasma immersed in superthermal radiation resembles a Lorentzian-type distribution.The Maxwellian and kappa distributions differ substantially in the high energy tail, but the differences become less significant as $\kappa$ increases.       

Although the kappa distribution has been known for several decades, it has been mostly used for the study of linear waves. Under linear approximation, kappa distribution employed to study the dispersive properties as well as the rate of Landau damping for a number of wave modes in space plasmas such as the electron plasma waves, ion acoustic waves, and electromagnetic R-mode and L-mode waves and noted that the presence of a high energy tail leads to a significant change in the damping rate of the interested wave compared to that of Maxwellian plasma \cite{6,7}. However, the linear approximation can break down because of various well-known instability mechanisms leading to the nonlinear regime. Thus, it provides a motivation for studying the nonlinear effect in a plasma medium that its equilibrium state is non-Maxwellian. Note that the plasma equilibrium always has essential influence on the nonlinear dynamics. Nonlinear effects either directly depends on the velocity distribution, such as particle trapping or is affected through the average of the distribution function. 

In this paper, the kappa distribution function is used in the study of the ion-acoustic soliton. The ion-acoustic solitons in a plasma with cold ions were initially investigated by Sagdeev \cite{8}. He assumed that the distribution of the electrons in the field of the wave has the equilibrium (Boltzmann) form. However, in the cases that particles trapping by the potential field of the wave is possible, the arrangement of particles in the phase space drastically changes and the closed trajectories are obtained. These particles cause a large deviation of distribution function in comparison to the case when the trapped particles are not considered. Here, we follow the well-established kinetic model of trapping in an adiabatically varying field \cite{9,10}. 

The present paper is devoted to a detailed calculations of the electron density from the kappa distribution function, including the effects of electron trapping. Since the density can not be expressed in terms of simple functions, therefore assuming a small potential energy (in comparison to the thermal energy), a Taylor expansion of the electron density is obtained. The electron density together with the fluid equations for the ions and the Poisson's equation form a complete set of equations that are used to study the ion-acoustic solitons. It is shown, as the most important result, that the deviation from Maxwellian distribution makes it possible to construct the localized structures which move subsonically. To our knowledge this result is quite new. Moreover, the soliton velocity and width decrease according to the measure of deviation.

The lay out of the paper is as follows. In Sec. II we give the basic mathematical calculation of the electron density. In Sec. III the complete formalism and the results are presented. Section IV is devoted to the summery and conclusions.

\section{The Electron Density}

Let us begin with the one dimensional kappa distribution function, $f_{\kappa}$ (Fig. 1), for a free system as follows: 
\begin{equation}\label{2}
f_{\kappa}(v)=\frac{n_0}{\sqrt{\pi}}\frac{1}{\theta}\frac{\Gamma(\kappa+1)}{\kappa^{3/2}\Gamma(\kappa-1/2)}
\left(1+\frac{v^2}{\kappa\theta^2}\right)^{-\kappa}, 
\end{equation}
where $\kappa$ is the spectral index; the thermal speed $\theta$ is related to the particle temperature $T$ by 
\begin{equation}\label{3}
\theta=[(2\kappa-3)/\kappa]^{1/2}(T/m)^{1/2}
\end{equation}
when $\kappa>3/2$; $\Gamma$ is the gamma function; and $f_\kappa$ has been normalized so that $\int f_\kappa dv=n_0$.
\begin{figure}
  \epsfxsize=8truecm
  \epsfysize=8truecm
\centerline{\epsfbox{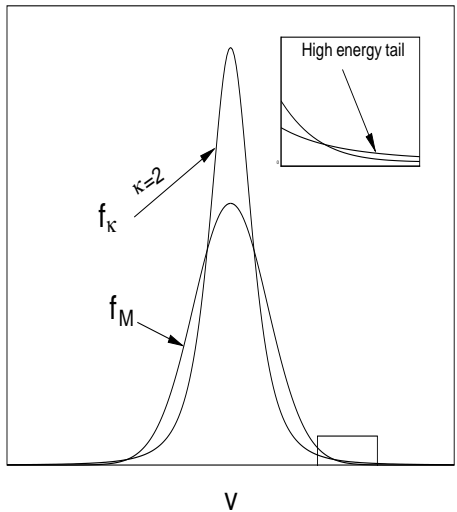}}
\caption{The Maxwellian and non-Maxwellian distributions. The small rectangular shows the high energy tail associated with the non-Maxwellian distribution.}
\label{fig:fig1}
\end{figure}
We note that as $\kappa\rightarrow\infty$, $f_{\kappa} \rightarrow f_M$, where $f_M$ is the Maxwellian distribution function given by
\begin{equation}\label{4}
f_{M}=\frac{n_{0}}{v_{T}\sqrt{2\pi }}\ \exp (-\frac{v^{2}}{2v_{T}^{2}}), 
\end{equation}
where $v_T^2=T/m$.

We need to define an upper limit for $\kappa$ above which $f_{\kappa}$ and $f_M$, up to the desirable accuracy, are almost the same. The width of the distribution functions can be considered as a convenient choice for this purpose.
We keep the error of the calculation of the order of one percent, and look for a solution of the following equation in terms of $\kappa$:
\begin{equation}\label{5}
\left| \frac{\Delta _{f_M}-\Delta f_{\kappa}}{\Delta _{f_M}}\right| \times 100=1, 
\end{equation}
where 
\begin{equation}\label{6}
\Delta _{f_M}=2\sqrt{2}v_{T}, 
\end{equation}
and
\begin{equation}\label{7}
\Delta f_{\kappa}=2\theta\sqrt{\kappa \left[\exp (1/\kappa)-1\right]},
\end{equation}
are the width of Maxwellian and non-Maxwellian distribution functions, respectively. The solution of the above equation results in $\kappa_{\max }=50$ for the maximum value of $\kappa$.

The above equilibriums are now extended to the case when the electrons confront a well-shape potential energy which modifies the distribution function in the following manner: 
\begin{equation}\label{8}
f_{\kappa}\left( v,v_{1}\right) =\frac{n_0}{\sqrt{\pi}}\frac{1}{\theta}\frac{\Gamma(\kappa+1)}{\kappa^{3/2}\Gamma(\kappa-1/2)}
\left(1+\frac{v^2-v_1^2}{\kappa\theta^2}\right)^{-\kappa}, 
\end{equation}
with the following definitions: 
\begin{eqnarray}
&&n_{0}=2\int_{0}^{\infty }f_{\kappa}\left( v,0\right) dv,\label{9}\\
&&v_{1}^{2}=2U/m, \label{10}
\end{eqnarray}
where $U$ is just the absolute value of the potential energy.

As it was mentioned earlier, in a collisionless plasma the distribution of the trapped electrons may differ significantly from the  equilibrium distribution, which influences the properties of the wave. The particles density, including the trapped particles, in a Maxwellian plasma in which the fields evolve adiabatically is \cite{9,10}:
\begin{equation}\label{11}
n(v_{1})=2\int_{v_{1}}^{\infty }f_M\left( v,v_{1}\right) dv+n_{0}\frac{2}{\sqrt{2\pi}}\frac{v_{1}}{v_T}.
\end{equation}
Accordingly, in the case of kappa distribution the modified number density becomes:
\begin{equation}\label{12}
n(v_{1})=2\int_{v_{1}}^{\infty }f_{\kappa}\left(v,v_1\right)dv+2
\frac{n_0}{\sqrt{\pi}}\frac{1}{\theta}\frac{\Gamma(\kappa+1)}{\kappa^{3/2}\Gamma(\kappa-1/2)}
v_{1}. 
\end{equation}
Since the above integral can not in general be evaluated in terms of simple functions (it can be generally expressed in term of hypergeometric functions), we expand $f_{\kappa}(v,v_1)$ for small $v_{1}(=\sqrt{2U/m})$ in terms of Taylor series and obtain
\begin{equation}\label{13}
\frac{n_e}{n_{0}}=1+\alpha _{\kappa}\left( \frac{U}{T}\right)
-\beta _{\kappa}\left( \frac{U}{T}\right) ^{3/2}+\gamma _{\kappa}\left( \frac{U}{T}%
\right) ^{2}, 
\end{equation}
where
\begin{eqnarray}
\alpha _{\kappa} &=&\frac{2\kappa-1}{2\kappa-3}, \label{14}\\
\beta _{\kappa} &=&\frac{4}{3\sqrt{\pi }}\frac{\Gamma \left( \kappa+1\right) }{\left(
\kappa-3/2\right) ^{5/2}\Gamma \left( \kappa-3/2\right) }, \label{15}\\
\gamma _{\kappa} &=&\frac{1}{2}\frac{\left( \kappa+1/2\right) \left( \kappa-1/2\right) }{%
\left( \kappa-3/2\right) ^{2}}. \label{16}
\end{eqnarray}

We note that we have retained, in the expansion, the term that is proportional to $U^{2}$. It is now possible to calculate the density of Eq. (\ref{12}) numerically and compare it with Eq. (\ref{13})(Fig. 1). 
\begin{figure}
  \epsfxsize=8truecm
  \epsfysize=8truecm
\centerline{\epsfbox{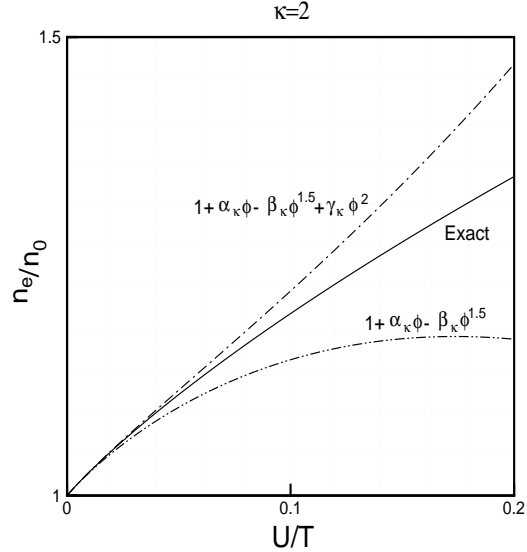}}
\caption{The normalized electron density versus the normalized potential energy.}
\label{fig:fig2}
\end{figure}
The deviation of the expansions from the exact density is clear. The expansion is down up to $U^{2}$ for the following reason. Quantitatively, we define the error of the expanded density in the following form,  
\begin{equation}\label{17}
\text{error}=\left| \frac{n_{e\text{ exact}}-n_{e\text{ expansion}}}{n_{e%
\text{ exact}}}\right| \times 100.
\end{equation}
Figure 3 depicts the error for different $\kappa$s versus the normalized potential energy. 
\begin{figure}
  \epsfxsize=8truecm
  \epsfysize=8truecm
\centerline{\epsfbox{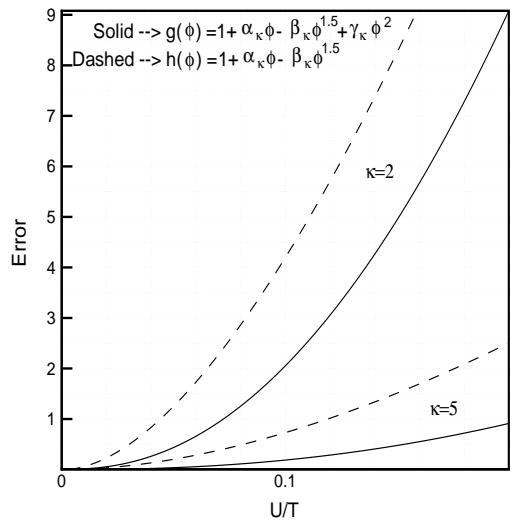}}
\caption{The error versus the normalized potential energy.}
\label{fig:fig3}
\end{figure}
It is clear that by increasing the $\kappa$ the error decreases. Therefore, in order to keep the error of the order of one percent, the $U^2$ term has to be kept in the density and the upper limit of the normalized potential energy should be restricted to 0.1.   

\section{Ion-acoustic soliton}

The ion motions are governed by the following standard equations for cold ions: 
\begin{eqnarray}
&&\frac{\partial n_i}{\partial t}+\frac{\partial}{\partial x}\left(n_i
v_i\right)=0 ,  \label{18} \\
&&\frac{\partial v_i}{\partial t}+v_i\frac{\partial}{\partial x} v_i=-\frac{%
\partial}{\partial x}\phi ,  \label{19} \\
&&\frac{\partial^2 \phi}{\partial x^2}=n_e-n_i,  \label{20}
\end{eqnarray}
where $n_e$ and $n_i$ are the electron and ion densities, $v_i$ is the ion velocity, and $\phi$ is the ambipolar potential due to charge separation that have been normalized with the quantities as follows: 
\begin{eqnarray}  \label{21}
&&\frac{n_i}{n_0}\rightarrow n_i,~~~~~\frac{n_e}{n_0}\rightarrow n_e,~~~~~%
\frac{v_i}{c_s}\rightarrow v_i,  \nonumber \\
&&\frac{x}{\lambda_D}\rightarrow x,~~~~~{\omega_{pi}}{t}\rightarrow
t,~~~~~~~~\frac{e\phi}{T}\rightarrow \phi.
\end{eqnarray}
Here $e$ is the magnitude of the electron charge, $n_0$ is the equilibrium density, $c_s(=\sqrt{T/m_i}$) is the ion-sound velocity, $\lambda_D[=\sqrt{T/(4 \pi e^2 n_0)}]$ is the Debye length, $\omega^2_{pi}=4\pi n_0e^2/m_i$, where $m_i$ and $T$ are the ion mass and electron temperature, respectively.

Now, assume the stationary case when all the quantities depend on $\xi =x-Mt$ where $M$ is the velocity of the stationary profiles. Localized solutions are considered, with the following conditions that as $\xi \rightarrow \infty$:
\begin{equation}
~~n_{e,i}\rightarrow 1,~~~~~\phi \ \text{and}\ d\phi /d\xi
\rightarrow 0,~~~~~v_{i}\rightarrow 0.  \label{22}
\end{equation}
Then Eq. (\ref{18}) and (\ref{19}) give 
\begin{equation}
n_{i}=\left( 1-\frac{2\phi }{M^{2}}\right) ^{-1/2}.  \label{23}
\end{equation}
It is worthwhile mentioning here that in order to avoid the hydrodynamic shock, the condition $\phi_{m} < M^2/2$ has to be fulfilled. For the $\phi_m \leq 0.1$, the results (Fig. 4) show that the above condition is satisfied over whole velocity range. Then, the ion density can be expanded for the shallow well ($\phi \ll 1$) as follows, 
\begin{equation}\label{24}
n_{i}=1+\frac{\phi }{M^{2}}+\frac{3}{2}\frac{\phi ^{2}}{M^{4}}.
\end{equation}

Substituting the electron density from Eq. (\ref{13}), ion density from Eq. (\ref{24}) into Eq. (\ref{20}) and integrating once will result to the following equation: 
\begin{eqnarray}
&&\frac{1}{2}\left( \frac{d\phi }{d\xi }\right) ^{2}+V(\phi )=0,
\label{26} \\
&&V(\phi )=-\frac{1}{2}\left( \alpha_\kappa -\frac{1}{M^{2}}\right) \phi^{2}+\frac{2}{5}\beta_\kappa \phi^{5/2}-
\frac{1}{3}\left( \gamma_\kappa-\frac{3}{2M^{4}}\right) \phi ^{3},  \label{27}
\end{eqnarray}
where the conditions (\ref{22}) are used to determine the constant of integration. Equation (\ref{27}) is the kappa counterpart of the Gurevich potential \cite{9} for $\phi \ll 1$. In fact, it is reduced to the Gurevich potential if one substitute the asymptotic values of $\alpha_\kappa=1$, $\beta_\kappa=4/(3\sqrt{\pi})$, and $\gamma_\kappa=1/2$ in Eqs. (\ref{14}), (\ref{15}), and (\ref{16}) when $\kappa \rightarrow \infty$.

\begin{figure}
  \epsfxsize=8truecm
  \epsfysize=8truecm
\centerline{\epsfbox{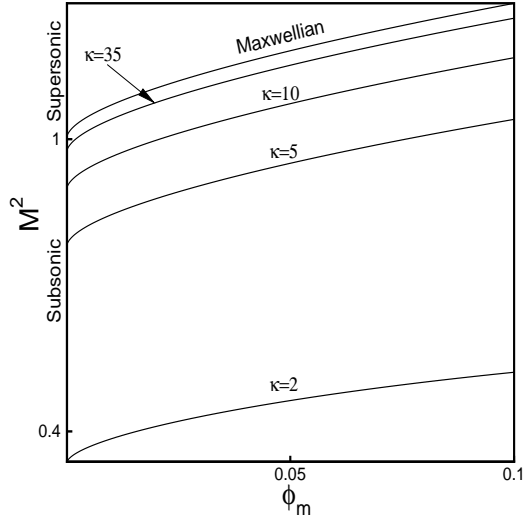}}
\caption{The ion-acoustic soliton velocity versus its maximum amplitude.}
\label{fig:fig4}
\end{figure}

For a soliton solution the effective potential of Eq. (\ref{27}) should fulfill certain conditions.  As it is seen, $V(\phi)$ and its first derivative vanish at $\phi=0$ and further, $V(\phi)$ has to be well-shaped between $0 \leq \phi \leq \phi_m$, where $\phi_m$ is the nonzero root of $V(\phi)=0$. That means, $\alpha_\kappa > 1/M^2$ and 
\begin{equation}\label{28}
M^2=\left[-\frac{1}{2\phi_m}+\sqrt{\frac{1}{4\phi_m^2}+\frac{\alpha_\kappa}{\phi_m}-\frac{4}{5}\frac{\beta_\kappa}{\sqrt{\phi_m}}+\frac{2}{3}\gamma_\kappa}\right]^{-1}.
\end{equation} 
 
Figures 4 and 5 show the behavior of $M^2$ versus $\kappa$ and $\phi_m$. As it is seen from Fig. 4, solitons with larger amplitude move faster. Therefore, since the largest amplitude is $0.1$, the upper limit of the soliton's velocity can be defined as follows:
\begin{equation}\label{29}
\frac{1}{\alpha_\kappa}<M^2<\left[-5+\sqrt{25+10\alpha_\kappa-\frac{4\sqrt{10}}{5}\beta_\kappa+\frac{2}{3}\gamma_\kappa}\right]^{-1}.
\end{equation}
\begin{figure}
  \epsfxsize=8truecm
  \epsfysize=8truecm
\centerline{\epsfbox{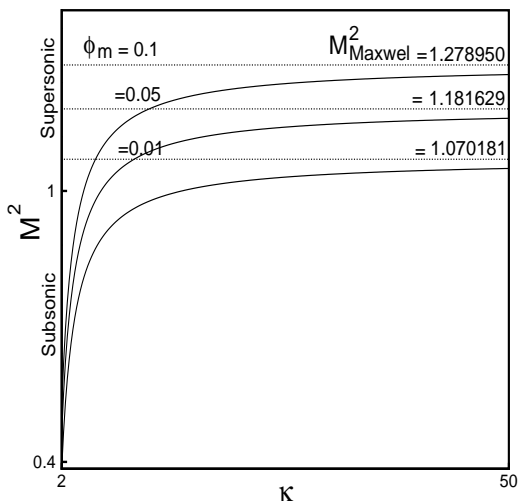}}
\caption{The ion-acoustic soliton velocity versus $\kappa$.}
\label{fig:fig5}
\end{figure}

It is very important to note that, the deviation form Maxwellian distribution (finite $\kappa$) makes it possible to construct ion-acoustic solitons in the subsonic regime (Figs. 4, 5, and 6). As it was mentioned earlier, the velocity of ion-acoustic solitons is a monotonically increasing function of its maximum amplitude. In the amplitude interval over which the above theory is based (0 $\leq \phi_m \leq 0.1$), as it is seen from Figs. 4 and 5, that solitons, resulting from small $\kappa$ ($\kappa<6$) are subsonic. Solitons associated with the larger $\kappa$, according to their maximum amplitude move sub or supersonically. Figure 6 shows the boundary of subsonic and supersonic regimes ($M=1$). 
\begin{figure}
  \epsfxsize=8truecm
  \epsfysize=8truecm
\centerline{\epsfbox{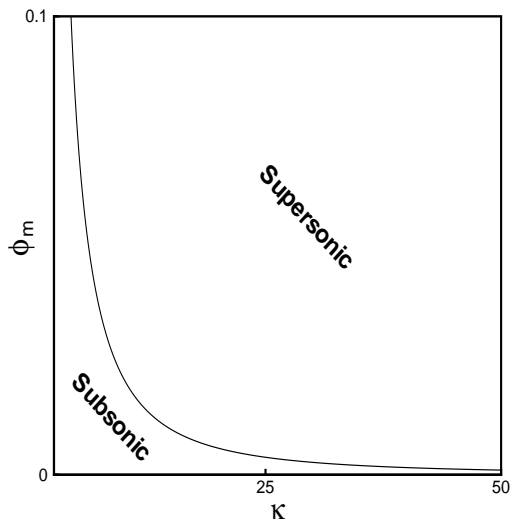}}
\caption{The sub and supersonic regimes.}
\label{fig:fig6}
\end{figure}
For larger $\kappa$ the amplitude interval over which the solitons are subsonic gets narrower. However, the existence of subsonic soliton is particular result of non-Maxwellian plasmas. Solitons move supersonically all over the amplitude interval just when $\kappa$ approaches to infinity, or in the other words, when the plasma electrons become Maxwellian.

We now discuss the width of the soliton. The width of the soliton is defined as
\begin{equation}\label{30}
\Delta=\int^{a_{max}/e^\prime}_{a_{max}}\frac{da}{\sqrt{-2V(a)}}, 
\end{equation}  
where $e^\prime$ in the upper limit is the Neperian number and the maximum 
amplitude has been assumed to be at $\xi=0$.

\begin{figure}
  \epsfxsize=8truecm
  \epsfysize=8truecm
\centerline{\epsfbox{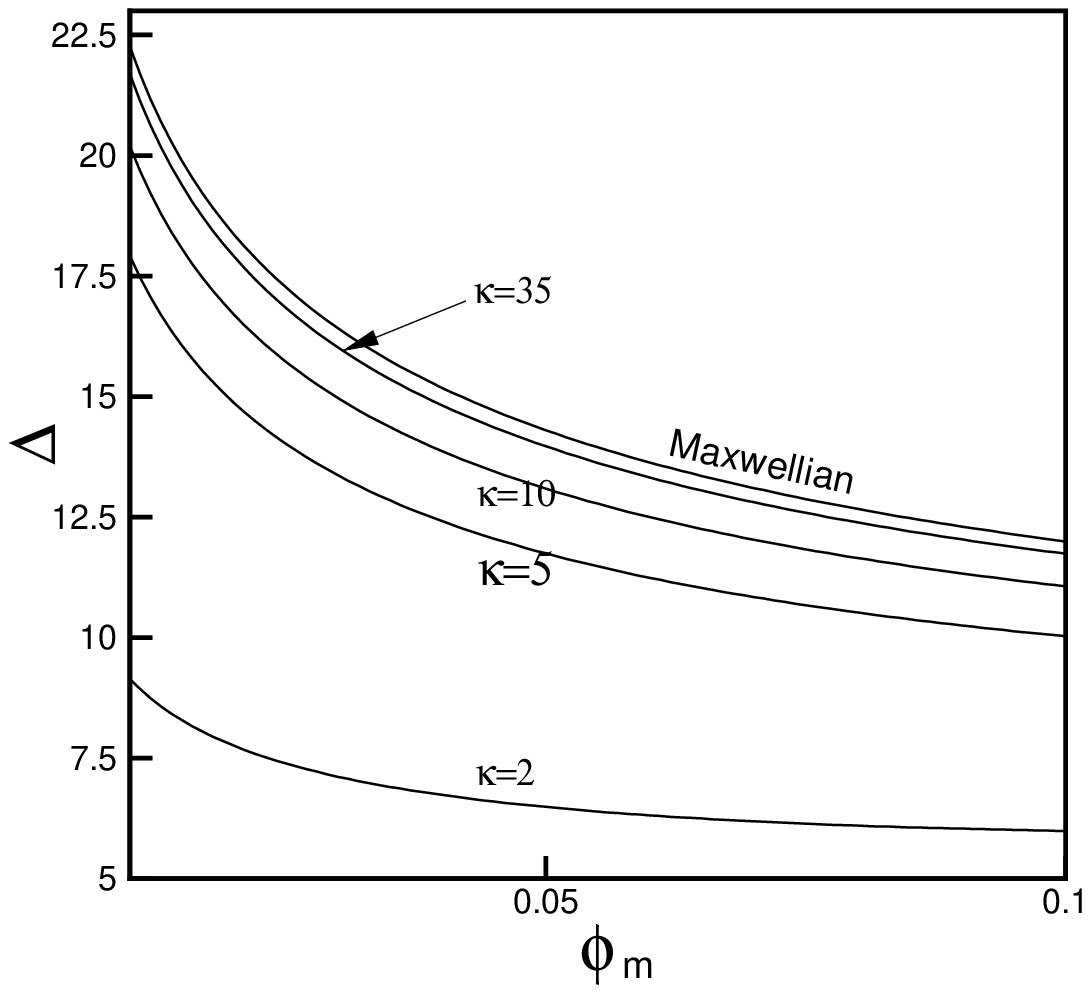}}
\caption{The sub and supersonic regimes.}
\label{fig:fig7}
\end{figure}

Figures 7 and 8 exhibit the dependence of the width $\Delta$ on the maximum amplitude $\phi_{m}$ and $\kappa$.
Apparently, for any $\phi_{m}$ the width of the solitons resulting from non-Maxwellian deviation 
is narrower according to the measure of its deviation from Maxwellian equilibrium.

\begin{figure}
  \epsfxsize=8truecm
  \epsfysize=8truecm
\centerline{\epsfbox{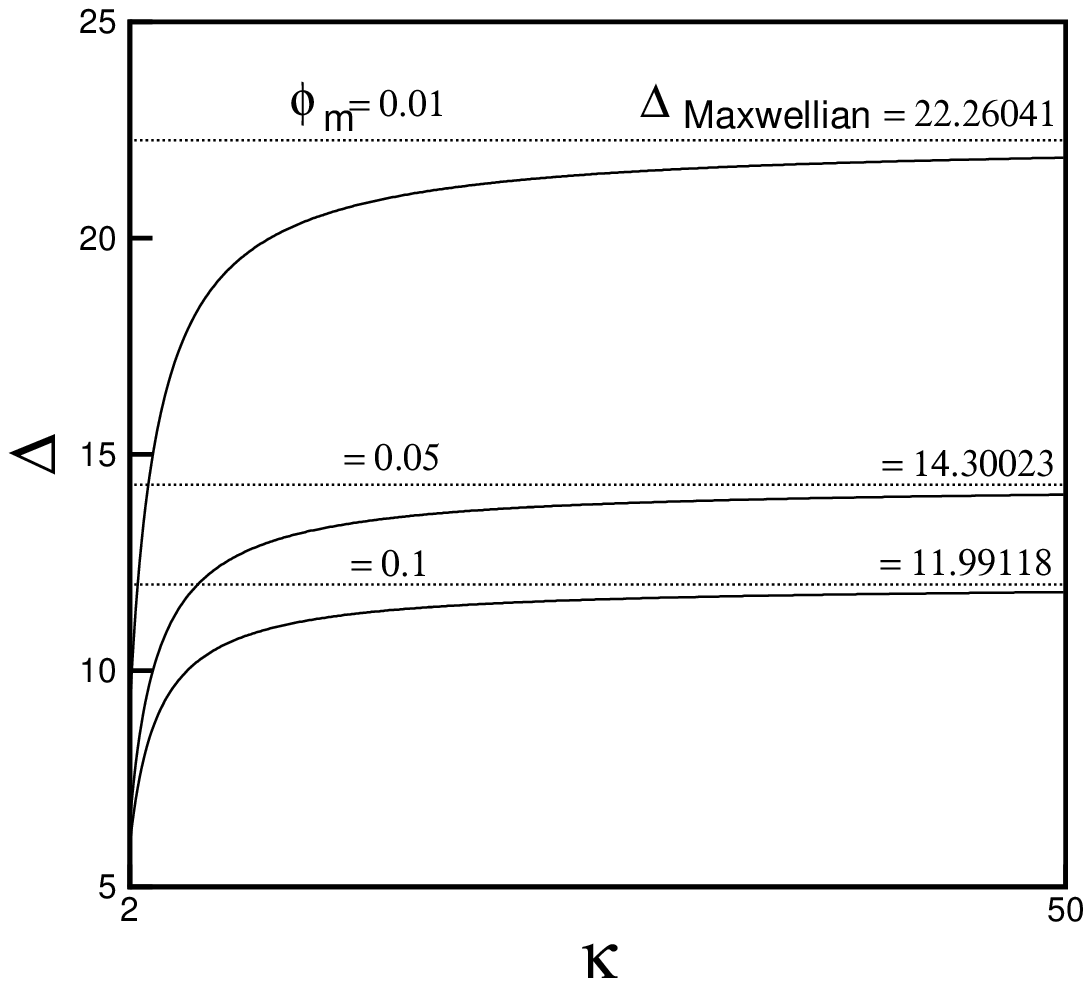}}
\caption{The sub and supersonic regimes.}
\label{fig:fig8}
\end{figure}

\section{summary and conclusions}
In this paper, the nonlinear theory of plasma waves was extended to the plasmas that their equilibrium state are specified by the non-Maxwellian (here kappa) distribution. We believe that the extension is very important since most of the space and some of laboratory plasmas are not in the Maxwellian equilibrium. Although the linear theory of this issue has been known for the decades but, to our knowledge, this is the first attempt in opening the gate to the nonlinear world of plasma waves in the non-Maxwellian background. As an example the ion-acoustic solitons are studied in this framework and it was shown that as the most important result regarding to the non-Maxwillian equilibrium is the possibility for solitons to move subsonically. The solitons velocity and width are monotonically increasing function of spectral index ($\kappa$) and approach to the their Maxwellian values as $\kappa \rightarrow \infty$.

%\end{multicols}

\end{document}